\documentstyle[prl,aps,epsf,epsfig,array,amsmath]{revtex}
\newcommand{\be}{\begin{equation}}
\newcommand{\ee}{\end{equation}}
\newcommand{\bqn}{\begin{eqnarray}}
\newcommand{\eqn}{\end{eqnarray}}
\begin{document}

\title{{\it Sum rules} in the oscillator radiation processes}
\author{R. Casana\thanks{E-mail: casana@ift.unesp.br}, G.
Flores-Hidalgo\thanks{E-mail: gflores@ift.unesp.br} and B. M.
Pimentel\thanks{E-mail: pimentel@ift.unesp.br}}
\address{ Instituto de F\'{\i}sica Te\'orica-IFT/UNESP, Rua Pamplona
145,\\ 01405-900, S\~ao Paulo, SP, Brazil}
\date{\today }
\maketitle
\begin{abstract}
We consider the problem of an harmonic oscillator coupled to a scalar field
in the framework of recently introduced dressed coordinates. We compute all the
probabilities associated with the decay process of an excited level of the oscillator.
Instead of doing direct quantum mechanical calculations we establish some {\it sum
rules} from which we infer the probabilities associated to the different decay
processes of the oscillator. Thus, the {\it sum rules} allows to show that the transition
probabilities between excited levels follow a binomial distribution.

\vspace{0.34cm}
\noindent PACS Number(s):~03.65.Ca, 32.80.Pj
\end{abstract}

\setlength{\topmargin}{-2ex}
\vskip2pc
\section{Introduction}
Recently, in analogy with the renormalized fields in quantum field theory, the
concept of {\it dressed coordinates} has been introduced \cite{adolfo1,adolfo2,gabriel}.
This concept was introduced in the context of an harmonic oscillator
interacting linearly with a massless scalar field and allowed the
construction of {\it dressed states}, with the physically correct
property of stability of the oscillator ground state in the absence of
field quanta (the quantum vacuum).  This concept also showed the technical
advantage of allowing an exact nonperturbative treatment of the oscillator
radiation processes.

Indeed, in spite of the system being linear, if we work with the usual
coordinates it is not possible an analytical computation of the
probabilities associated with the oscillator radiation processes. For
example, the exact calculation of the simplest process, the probability
for the oscillator to decay spontaneously from the first excited level
to the ground state is not possible. However in the framework of
dressed coordinates, it has been possible to solve this problem exactly
and for small coupling constant, it agrees with the long time know
result: $e^{-\Gamma t}$ \cite{adolfo1}. Furthermore, when applied to a
confined atom (approximated by the oscillator) in a spherical cavity of
sufficiently small radius \cite{adolfo2}, the method accounted for, the
experimentally observed, inhibition of the spontaneous decaying
processes \cite{hulet,haroche2}. Also, in Refs. \cite{nonlinear,yony}
the case of the nonlinear interaction between the oscillator and the
field modes was treated.

Nevertheless, in all previous works \cite{adolfo1,adolfo2,gabriel} it
was only computed the probability associated with the simplest
process just above described. The aim of this paper is to
fill this gap. In doing this, we introduce a calculational method that we
believe could be extended to other physical situations. Instead  of performing
direct computations of the quantum mechanical formulae we establish
some identities, that we named as {\it sum rules}, and from them
we infer the probabilities associated with the different oscillator
radiation processes.

The paper is organized as follows. In section II we review the
concept of dressed coordinates and dressed states. Section III is
devoted to the direct computation of the probabilities associated with
simplest decaying processes of the oscillator. In section IV we introduce the
{\it sum rules}. Finally, in section V we give our concluding remarks.
Along this paper we use natural units $c=\hbar =1$.

\section{Dressed Coordinates and Dressed States}
In this section, in order to make this paper self contained, we review the concept of
dressed coordinates and dressed states as defined in Refs. \cite{adolfo1,adolfo2,gabriel}.
For this end we consider the system composed by an harmonic
oscillator linearly coupled to a massless scalar field. The Hamiltonian for
this system is given by \cite{adolfo1}

\begin{equation}
H=\frac{1}{2}\left( p_{0}^{2}+\omega _{0}^{2}q_{0}^{2}\right)
+\frac{1}{2} \sum_{k=1}^{N}\left( p_{k}^{2}+\omega _{k}^{2}
q_{k}^{2}\right) -\sum_{k=1}^{N}c_{k}q_{k}q_0+\frac{1}{2}
\sum_{k=1}^{N}\frac{c_{k}^{2}} { \omega _{k}^{2}}q_{0}^{2}\;,
\label{Ham}
\end{equation}
where $q_{0}$ is the oscillator coordinate, $q_{k}$ are the field
modes; $k=1,2,...$; $\omega _{k}=2\pi /L$, $c_{k}=\eta \omega
_{k}$, $\eta = \sqrt{2g\Delta \omega }$, $\Delta \omega =\omega
_{k+1}-\omega _{k}= 2\pi /L$, $g$ is a frequency dimensional
coupling constant and $L$ the diameter of the sphere in which we
confine the oscillator-field system. In Eq. (\ref{Ham}) the limit
$N\rightarrow \infty $ is understood. The last term in Eq. (\ref{Ham}) guarantees
the  positiveness of the Hamiltonian and it can be
seen as a renormalization of the oscillator frequency \cite{tirring,weiss}.

To understand the necessity of introducing  dressed coordinates and dressed
states let us
consider the following problem. Imagine that the oscillator of frequency
$\omega_0$ in Eq. (\ref{Ham}) represents an atom and the other oscillators represent
the modes of the electromagnetic field. If there are no interactions among
them the free Hamiltonian, $c_k=0$, has the following eigenfunctions,

\bqn
\psi_{n_0n_1...n_N}(q)&\equiv&
\langle q|n_0,n_1,...,n_N\rangle
\nonumber\\
&=&\prod_{\mu=0}^N\left[\left(\frac{\omega_\mu}{\pi}\right)^{1/4}
\sqrt{\frac{2^{-n_\mu}}{n_\mu!}}H_{n_{\mu}}(\sqrt{\omega_{\mu}}q_\mu)
e^{-\frac{1}{2}\omega_\mu q_\mu^2}\right]\;.
\label{e8a}
\eqn
The physical meaning of $\psi_{n_0n_1...n_N}(q)$ in this case is clear,
it represents the atom in its $n_0$-th excited level and $n_k$ photons
of frequencies $\omega_k$. Now, consider the state
$\psi_{n_00...0}(q)$: the excited atom in the quantum vacuum. We know
from experience that any excited level of the atom is unstable. The
explanation of this fact is that the atom is not isolated from
interacting with the quantum electromagnetic field. This interaction in
our toy model is given by the linear coupling of $q_0$ with $q_k$.
Obviously, when we take into account this interaction any state of the
type $\psi_{n_00...0}(q)$ is rendered unstable. But, there is a
problem, the state $\psi_{00...0}(q)$, that represents the atom in its
ground state and no photons, is also unstable contradicting the
experimental fact of the stability of the atom ground state. What is
wrong? The first thing that cames in our mind is to think that the
model given by Eq. (\ref{Ham}) is wrong. Certainly, we know that the
correct theory to describe this physical system is quantum
electrodynamics. On the other hand such a description could be
extremely complicated. If we aim to  maintain the model as simple as
possible and still insist in describing it by the Hamiltonian given in
Eq. (\ref{Ham}) what we can do in order to take into account the
stability of the atom ground state? The answer lies in the spirit of
the renormalization program in quantum field theory: the coordinates
$q_\mu$ that appear in the Hamiltonian  are not the physical ones, they
are bare coordinates. We introduce dressed (or renormalized)
coordinates, $q_0'$ and $q_k'$, respectively for the dressed atom and
the dressed photons. We define these coordinates as the physically
meaningful ones. In terms of these coordinates we define the dressed
states by

\bqn
\psi_{n_0n_1...n_N}(q')&\equiv&
\langle q'|n_0,n_1,...,n_N\rangle_d \nonumber\\
&=&\prod_{\mu=0}^N\left[\left(\frac{\omega_\mu}{\pi}\right)^{1/4}
\sqrt{\frac{2^{-n_\mu}}{n_\mu!}}H_{n_{\mu}}(\sqrt{\omega_{\mu}}q_\mu')
e^{-\frac{1}{2}\omega_\mu (q_\mu')^2}\right]
\label{e8b}
\eqn
where the subscript $d$ means dressed state. The dressed states given by Eq.
(\ref{e8b}) are defined as the physically measurable
states and describe in general, the physical atom in the
$n_0$-th excited level and $n_k$ physical photons of frequency $\omega_k$.
Obviously, in the limit in which the coupling constant $c_k$
vanishes the renormalized coordinates $q_\mu'$ must approach the bare coordinates $q_\mu$.
Now, in order to relate the bare and dressed coordinates we have to use the physical
requirement of stability of the dressed ground state. The dressed ground
state will be stable only and only if it is defined as the eigenfunction of the
interacting Hamiltonian given by Eq. (\ref{Ham}). Also the dressed ground state
must be the one of minimum energy, that is, it must be defined as being
identical (or proportional) to the ground state eigenfunction of the
interacting Hamiltonian. From this definition, one can construct the dressed
coordinates in terms of the bare ones.  Then, the first step in order
to obtain the dressed coordinates is to solve the eigenfunctions of
the Hamiltonian given in Eq. (\ref{Ham}). This bilinear Hamiltonian can be diagonalized
by introduzing normal coordinates and momenta $Q_r$ and $P_r$,
\begin{equation}
q_{\mu }=\sum_{r=0}^{N}t_{\mu }^{r}Q_{r}\;,~~~p_{\mu }=\sum_{r=0}^{N}
t_{\mu}^{r}P_{r}\;,~~~\mu =(0,k)\;,
\label{transf}
\end{equation}
where $\{t_{\mu }^{r}\}$ is an orthonormal matrix whose elements are given by
\cite{gabrielrudnei},
\begin{equation}
t_{k}^{r}=\frac{c_{k}}{(\omega_{k}^{2}-\Omega_{r}^{2})}t_{0}^{r}\;,
~~~~~~~~~~~ t_{0}^{r}= \left[1+\sum_{k=1}^{N}\frac{c_{k}^{2}}
{(\omega_{k}^{2}-\Omega_{r}^{2})^{2}}\right]^{-\frac{1}{2}}
\label{tkrg1}
\end{equation}
with $\Omega_r$ being the normal frequencies corresponding to the
collective modes of the coupled system and given as solutions of the
equation
\begin{equation}
\omega_0^2-\Omega_r^2=\sum_{k=1}^N\frac{c_k^2\Omega_r^2}
{\omega_k^2(\omega_k^2-\Omega_r^2)}\;.
\label{la5b}
\end{equation}
In terms of normal coordinates and momenta the Hamiltonian given by Eq. (\ref{Ham})
reads as
\begin{equation}
H=\frac{1}{2}\sum_{r=0}^{N}(P_{r}^{2}+\Omega_{r}^{2}Q_{r}^{2})\;,
\label{diagonal}
\end{equation}
then, the eigenfunctions of the Hamiltonian can be written as
\begin{eqnarray}
\phi_{n_{0}n_{1}...n_N}(Q)&\equiv &
\langle Q|n_{0},n_{1},...,n_N\rangle_c
\nonumber \\
&=&\prod_{r=0}^N\left[\left(\frac{\Omega_r}{\pi}\right)^{1/4}
\sqrt{\frac{2^{-n_r}}{n_r!}}H_{n_{r}}(\sqrt{\Omega_{r}} Q_{r})
e^{-\frac{1}{2}\Omega_r^2}\right]\;,
\label{autofuncoes}
\end{eqnarray}
where the subscript $c$ means collective state. Now, using the definition
of the dressed coordinates: $\psi_{00...0}(q')\propto\phi_{00...0}(Q)$,
and using Eqs. (\ref{e8b}) and (\ref{autofuncoes}) we get
$e^{-\frac{1}{2}\sum_{\mu=0}^N\omega_\mu (q_\mu')^2}=
e^{-\frac{1}{2}\sum_{r=0}^N\Omega_r Q_r^2}$, from which
the dressed coordinates are obtained
\begin{equation}
q_\mu'=\sum_{r=0}^N\sqrt{\frac{\Omega_r}{\omega_\mu}}t_\mu^rQ_r\;.
\label{dress}
\end{equation}

We have to point out that the dressed coordinates here introduced are not
simply a change of variables, they are new coordinates in its own right and
are introduced by physical consistence requirement of the model. Also we have
to stress that our dressed coordinates are not the same as the ones employed in
other references, as
for example in \cite{prigogine} and references therein, where the authors called
dressed coordinates the collective normal ones. Also our dressed states
are  different from the ones defined in Refs.
\cite{polonsky,haroche,haroche1,cohen,cohen1}, where the authors called dressed
states the states obtained by diagonalizing the Hamiltonian of a two level
atom coupled to an finite number of electromagnetic field modes.

Before leaving this section we would like to remark that a similar
model to the one given by Eq. (\ref{Ham}) has been used repeatedly from
time to time as a simplified model to describe the quantum Brownian
motion \cite{feynman,ulersma,caldeira,zurek1}, the decoherence problem
and other related problems \cite{zurek2}. Since the focus of these
works is other the one we have posed here, it is not necessary the
introduction of the dressed coordinates. On the other hand in quantum
optics, the model obtained from Eq. (\ref{Ham}) by considering the
rotating wave approximation has been extensively used
\cite{cohen,mandel,walls,davidovich}. For example, a common situation
that is described with this model is the problem of a cavity mode
(described by the oscillator of frequency $\omega_0$) coupled to the
external modes of the cavity. In this case, the introduction of the
dressed coordinates is unnecessary. The reason for this is that, in the
rotating wave approximation the linear interaction contains only terms
like $\hat{a}_0^\dag\hat{a}_k$ and $\hat{a}_k^\dag\hat{a}_0$ and since
$\hat{a}_0^\dag\hat{a}_k|0,0,...,0\rangle=0$ and
$\hat{a}_k^\dag\hat{a}_0|0,0,...,0\rangle=0$ it is guaranteed
automatically that ground state of the non interacting Hamiltonian is
also eigenstate of the interacting Hamiltonian, assuring in this way
the stability of the state with no photons both inside and outside of
the cavity. Of course if we use the model given by Eq. (\ref{Ham}),
with no rotating wave approximation, to describe the mentioned physical
situation it will be necessary the introduction of the renormalized
coordinates, otherwise the state of no photons both inside and outside
the cavity will evolve to an state of non zero photons in contradiction
with experiment. The advantage of our approach in treating this problem
will be the avoiding of the rotating wave approximation and also, from
a technical point of view the calculations will be greatly simplified.
In particular, this will be seen in next sections where we compute
easily the probabilities associated with the different oscillator
radiation processes.

\section{The decay processes}
We are interested mainly in the computation of the probabilities
associated with the different radiation processes of an excited state
of the oscillator. Thus, wishing to maintain the reasoning as general
as it is possible we show the necessary steps to compute the
probability amplitude associated with the most general transition:
Let the initial state of the system, at $t=0$, given by
$|n_0,n_1,...,n_N \rangle _{d}$, then, we ask what is the
probability amplitude of finding it at time $t$ in the state
$|m_0,m_1,...m_N\rangle _{d}$? Such probability amplitude, which
we denote as being ${\cal A}_{n_0n_1...n_N}^{m_0 m_1...m_N}(t)$, is given by
\begin{eqnarray}
{\cal A}_{n_0n_1...n_N}^{m_0m_1...m_N}(t) &=&~_d\langle m_0,m_1,...,m_N|
e^{-iHt}|n_0,n_1,...,n_N\rangle_{d}\nonumber\\
&=&\sum_{l_0l_1...l_N=0}^{\infty}T_{n_0n_1...n_N}^{l_0l_1...l_N}
T_{m_0m_1...m_N}^{l_0l_1...l_N}e^{-itE_{l_0l_1...l_N}}\;,
\label{dc1}
\end{eqnarray}
where $E_{l_0l_1...l_N}=\sum_{r=0}^N (l_r+\frac{1}{2})\Omega_r$ are the collective
energy eigenvalues and
\begin{eqnarray}
T_{n_0n_1...n_N}^{l_0l_1...l_N} &=&~_c\langle l_0,l_1,...,l_N
|n_0,n_1,...,n_N\rangle _{d}  \nonumber \\
&=&\int dQ\phi _{l_0l_1...l_N}(Q)\psi _{n_0n_1...n_N}(q^{\prime })
\label{dc2}
\end{eqnarray}
Above, the wave functions $\phi _{l_0l_1...l_N}(Q)$ and
$\psi_{n_0n_1...n_N}(q^{\prime })$ are normalized in $Q$ coordinates.
The wave function $\phi _{l_0l_1...l_N}(Q)$ as given in Eq.
(\ref{autofuncoes}) is already normalized. On the other hand the
dressed wave function $\psi_{n_0n_1...n_N} (q^{\prime })$ as given
in Eq. (\ref{e8b}) is normalized in the $q^{\prime }$ coordinates
but not in $Q$ coordinates. It is easy to show that if the dressed
ground state $\psi_{00...0}(q')$ is normalized in $Q$ coordinates then,
automatically all the dressed states, given by Eq. (\ref{e8b}), are also
normalized.  Therefore, assuming that it is the case, we can replace Eqs.
(\ref{e8b}) and (\ref{autofuncoes}) in Eq. (\ref{dc2}) and using Eq.
(\ref{dress}) we get
\begin{equation}
T_{n_0n_1...n_N}^{l_0l_1...l_N}=\int dQ
\left[\prod_{r,\mu=0}^{N}\left(\frac{\Omega_r}{\pi}\right)^{1/2}
\sqrt{\frac{ 2^{-l_{r}}}{l_{r}!}}
\sqrt{\frac{2^{-n_{\mu }}}{n_{\mu}!}}
H_{l_{r}}( \sqrt{\Omega _{r}}Q_{r})
H_{n_{\mu}}(\sum_{s=0}^{N}t_{\mu }^{s}\sqrt{\Omega _{s}}Q_{s})\right]
e^{-\sum_{r=0}^N\Omega_r^2Q_r^2}\;.
\label{tcof}
\end{equation}
To compute the above integral it will be useful the following identity
\cite{bateman3},
\begin{equation}
H_{n}(\sum_{r=0}^Nt_{\mu}^{r}\sqrt{\Omega_{r}}Q_{r}) =n!
\sum_{s_0+s_1+...+s_N=n}\frac{(t_\mu^0)^{s_0}}{s_0!}
\frac{(t_\mu^1)^{s_1}}{ s_1!}...\frac{(t_\mu^N)^{s_N}}{s_N!}
H_{s_0}(\sqrt{\Omega_0}Q_0)H_{s_1}( \sqrt{\Omega_1}Q_1)...
H_{s_N}(\sqrt{\Omega_N}Q_N)\;.
\label{thed}
\end{equation}

Next, we compute the probability amplitude of an initial
state $|00...n_\mu...0\rangle_d$ in $t=0$ to be found after a time
$t$ in the state $|00... m_\nu...0\rangle_d$. From Eq. (\ref{dc1})
we get
\begin{equation}
{\cal A}_{00...n_\mu...0}^{00...m_\nu...0}(t)=
\sum_{l_0l_1...l_N=0}^\infty T^{l_0l_1...l_N}_{00...n_\mu...0}~\!
T^{l_0l_1...l_N}_{00...m_\nu...0}~\!
e^{-itE_{l_0l_1...l_N}}\;.
\label{z1}
\end{equation}
And after using Eq. (\ref{thed}) in (\ref{tcof}) we obtain
\begin{eqnarray}
T^{l_0l_1...l_N}_{00...n_\mu...0}=
\sqrt{n_\mu!} \sum_{s_0+s_1+...s_N= n_\mu}
\frac{(t_\mu^0)^{s_0}} {\sqrt{  s_0!}}\frac{(t_\mu^1)^{s_1}}{\sqrt{s_1!}}
...\frac{(t_\mu^N)^{s_N}}{\sqrt{s_N!}  }
\delta_{l_0s_0} \delta_{l_1s_1}...\delta_{l_Ns_N}\;.
\label{cof2}
\end{eqnarray}
Substituting Eq. (\ref{cof2}) in Eq. (\ref{z1}) we get
\begin{eqnarray}
{\cal A}_{00...n_\mu...0}^{00...m_\nu...0}(t)= e^{-itE_{00...0}}
\delta_{n m}\left[f_{\mu\nu}(t)\right]^n\;,
\label{z2}
\end{eqnarray}
where $f_{\mu\nu}(t)$ is given by
\begin{equation}
f_{\mu\nu}(t)=\sum_{r=0}^N t_\mu^r t_\nu^r e^{-i\Omega_r t}\;.
\label{z3}
\end{equation}

By setting $\mu=\nu=0$ in Eq. (\ref{z2}) we obtain the probability
amplitude for the oscillator to remain at time $t$ in the $n$-th
excited state, thus, it reads
\begin{equation}
{\cal A}_{n00...0}^{n00...0}(t)=\left[f_{00}(t)\right]^n\;,
\label{z4}
\end{equation}
where we have discarded the phase factor $e^{-itE_{000...}}$ because
it does not contribute for the associated probability. Also, by
setting $\mu=0$ and $\nu=k$ we can obtain the probability amplitude
of the particle to decay from the $n$-th excited level to its ground
state by emission of $n$ field quanta of frequencies (or energy in
$\hbar=1$ units) $\omega_k$,
\begin{equation}
{\cal A}_{n00...0}^{00...n_k...0}=\left[f_{0k}(t)\right]^n\;.
\label{z5}
\end{equation}
In the continuum limit, $L\to\infty$, the quantity $f_{00}(t)$, obtained from
Eq. (\ref{z3}) by setting $\mu=\nu=0$, has been computed in Ref. \cite{adolfo1}.
It reads as
\begin{equation}
f_{00}(t)=2g\int_{0}^\infty d\Omega\frac{\Omega^2 e^{-i\Omega t}}
{(\Omega^2-\omega_0^2)^2+\pi^2 g^2\Omega^2}\;.
\label{large}
\end{equation}
For $g/\omega_0\ll 1$, that corresponds to weak coupling, the integrand in above
equation is sharply peaked around $\Omega=\omega_0$ and in this case we can
obtain easily $f_{00}(t)=e^{-i\omega_o t-\Gamma t/2}$, with $\Gamma=\pi g$. Replacing this
result in Eq. (\ref{z4}) and taking the square modulus we obtain
for the probability that the particle oscillator remains in the $n$-th excited level, the
old know result: $e^{-n\Gamma t}$ \cite{expodecay}.

\section{\it sum rules}
It is clear that  if the oscillator is initially in its $n$-th
excited level it can decay of many different ways from that described
in Eq. (\ref{z5}). For example it can decay to its ground state
by emission of $n$ field quanta of different frequencies or it can
also decay to other lower excited states by emission of a number of
field quanta less than $n$. The probability amplitudes related to
these processes can all be computed by using Eqs. (\ref{dc1}) and (\ref{tcof}).
The task to be made can be very hard because it would be necessary to compute
the integral given in Eq. (\ref{tcof}) that involves products of more than two
Hermite polynomials, as it can be noted by substituting Eq. (\ref{thed}) in
(\ref{tcof}). Therefore, we will avoid this complication following an
alternative way. If the particle oscillator is at time $t=0$
in the $n$-th excited state we expect that at time $t$: it remains excited,
emit a field quantum of frequency $\omega_{k_1}$ and go to the $(n-1)$-th excited level,
emit two field quanta of frequencies $\omega_{k_1}$, $\omega_{k_2}$ and go to the
$(n-2)$-th excited level, emit three field quanta of frequencies $\omega_{k_1}$,
$\omega_{k_2}$, $\omega_{k_3}$ and go to the $(n-3)$-th excited level and so on.
We denote the probability amplitudes related with these processes respectively by
${\cal A}_{n00...0}^{(n-1)1_{k_1}}(t)$,
${\cal A}_{n00...0}^{(n-2)1_{k_1}1_{k_2}}(t)$,
${\cal A}_{n00...0}^{(n-3)1_{k_1}1_{k_2}1_{k_3}}(t)$, or in general
${\cal A}_{n00...0}^{(n-i)1_{k_1}1_{k_2}...1_{k_i}}(t)$, $i=1,2,...n$. The
corresponding probabilities are denoted by
${\cal P}_{n00...0}^{(n-1)1_{k_1}1_{k_2}...1_{k_i}}(t)$, $i=1,2...n$.
Now we will compute all these quantities from the knowledge of the probability
of the particle to remain excited in the $n$-th excited level,
${\cal P}_{n00...0}^{n00...0}(t)$ whose probability amplitude is given by Eq.
(\ref{z4}). For this end we use the identity
\begin{equation}
{\cal P}_{n00...0}^{n00...0}(t)+\sum_{k_1}{\cal P}_{n00...0}^{(n-1)1_{k_1}}(t)
+\sum_{k_1k_2}{\cal P}_{n00...0}^{(n-2)1_{k_1}1_{k_2}}(t)+...+
\sum_{k_1k_2...k_n}{\cal P}_{n00...0}^{01_{k_1}1_{k_2}...1_{k_n}}(t)=1\;,
\label{z6}
\end{equation}
that expresses, the sum of the probabilities of all possibilities is
equal to one. Starting from Eq. (\ref{z4}) we have to be able to get an identity
similar to Eq. (\ref{z6}) and from such expression identify the respective
probabilities associated with all the other possible processes.

From Eq. (\ref{z4}) and using Eq. (\ref{z3}) with $\mu=\nu=0$ we can
write  ${\cal P}_{n00...0}^{n00...0}(t)$ as
\begin{eqnarray}
{\cal P}_{n00...0}^{n00...0}(t)&=&\left(\sum_{rs}(t_0^r)^2(t_0^s)^2
e^{-it(\Omega_r-\Omega_s)}\right)^n  \nonumber \\
&=&\sum_{r_1s_1r_2s_2...r_ns_n}(t_0^{r_1}t_0^{s_1})^2(t_0^{r_2}
t_0^{s_2})^2...(t_0^{r_n}t_0^{s_n})^2e^{-it(\Omega_{r_1}+\Omega_{r_2}
+...+\Omega_{r_n}- \Omega_{s_1}-\Omega_{s_2}-...-\Omega_{s_n})}\;.
\label{z7}
\end{eqnarray}
On the other hand, by using the identity $\sum_{r}(t_0^r)^2=1$ we get
\begin{equation}
1=\sum_{r_1s_1r_2s_2...r_ns_n}t_0^{r_1}t_0^{s_1}t_0^{r_2}t_0^{s_2}...
t_0^{r_n}t_0^{s_n}\delta_{r_1s_1}\delta_{r_2s_2}...\delta_{r_ns_n}
e^{-it(\Omega_{r_1}+\Omega_{r_2}+...+\Omega_{r_n}-
\Omega_{s_1}-\Omega_{s_2}-...-\Omega_{s_n})}\;.
\label{z8}
\end{equation}
By adding and subtracting $1$ on the right hand side of the Eq.
(\ref{z7}) and using the relation given by Eq. (\ref{z8}), we obtain
\begin{eqnarray}
{\cal P}_{n00...0}^{n00...0}(t)&=&1-\sum_{r_1s_1r_2s_2...r_ns_n}
t_0^{r_1}t_0^{s_1}t_0^{r_2}t_0^{s_2}...t_0^{r_n}t_0^{s_n}
\left(\delta_{r_1s_1}\delta_{r_2s_2}...\delta_{r_ns_n}-
t_0^{r_1}t_0^{s_1}t_0^{r_2}t_0^{s_2}...t_0^{r_n}t_0^{s_n}\right)
\nonumber
\\
& &~~~~~~~~~~~~~~~~~~~~~\times
e^{-it(\Omega_{r_1}+\Omega_{r_2}+...+\Omega_{r_n}-
\Omega_{s_1}-\Omega_{s_2}-...-\Omega_{s_n})}\;.
\label{z9}
\end{eqnarray}
From the orthonormality of the matrix elements $\{t_\mu^r\}$ we have
$\sum_{\mu}t_\mu^rt_\mu^s=t_0^rt_0^s+\sum_k t_k^rt_k^s=\delta_{rs}$
and using it in Eq. (\ref{z9}) we get
\begin{eqnarray}
{\cal P}_{n00...0}^{n00...0}(t)&=& 1-\sum_{r_1s_1r_2s_2...r_ns_n}
t_0^{r_1}t_0^{s_1}t_0^{r_2}t_0^{s_2}...t_0^{r_n}t_0^{s_n}
\left[(t_0^{r_1}t_0^{s_1}+\sum_{k_1}t_{k_1}^{r_1}t_{k_1}^{s_1})
(t_0^{r_2}t_0^{s_2}+\sum_{k_2}t_{k_2}^{r_2}t_{k_2}^{s_2})...\right.
\nonumber \\
& &~~~~~~\left. ...(t_0^{r_n}t_0^{s_n}+\sum_{k_n}t_{k_n}^{r_n}
t_{k_n}^{s_n}) -t_0^{r_1}t_0^{s_1}t_0^{r_2}t_0^{s_2}...t_0^{r_n}
t_0^{s_n}\right] e^{-it(\Omega_{r_1}+\Omega_{r_2}+...+
\Omega_{r_n}-\Omega_{s_1}-\Omega_{s_2}-...-\Omega_{s_n})}  \nonumber \\
&=&1-\sum_{r_1s_1r_2s_2...r_ns_n}t_0^{r_1}t_0^{s_1}t_0^{r_2}t_0^{s_2}
... t_0^{r_n}t_0^{s_n} \left[\left(
\begin{array}{c}
n \\1\end{array} \right)
\sum_{k_1}t_{k_1}^{r_1}t_{k_1}^{s_1}t_0^{r_2}t_0^{s_2}...t_0^{r_n}
t_0^{s_n} \right.  \nonumber \\
& &~~~~~~~~~~~~~~~~~\left. +\left(
\begin{array}{c}
n \\
2  \end{array} \right)
\sum_{k_1k_2}t_{k_1}^{r_1}t_{k_1}^{s_1}t_{k_2}^{r_2}t_{k_2}^{s_2}
t_0^{r_3}t_0^{s_3}...t_0^{r_n}t_0^{s_n}+...
+\sum_{k_1k_2...k_n}t_{k_1}^{r_1}t_{k_1}^{s_1}t_{k_2}^{r_2}
t_{k_2}^{s_2}...t_{k_n}^{r_n}t_{k_n}^{s_n}\right]  \nonumber \\
& &~~~~~~~~~~~~~~~~~~~~~~~~\times
e^{-it(\Omega_{r_1}+\Omega_{r_2}+...+\Omega_{r_n}-\Omega_{s_1}
-\Omega_{s_2}-...-\Omega_{s_n})}\;,
\label{z10}
\end{eqnarray}
where in the second line we have used the symmetry of the expression
under index permutations. In terms of $f_{00}(t)$ and $f_{0k}$,
that can be obtained from Eq. (\ref{z3}), we can
write Eq. (\ref{z10}) as
\begin{eqnarray}
&&{\cal P}_{n0...0}^{n0...0}(t)+ \sum_{k_1}\left(
\begin{array}{c} n \\ 1  \end{array} \right) \left|f_{0k_1}
[f_{00}(t)]^{n-1}\right|^2+\sum_{k_1k_2}\left(\begin{array}{c} n
\\ 2 \end{array} \right)\left|f_{0k_1}(t)f_{0k_2}(t)[f_{00}(t)]^{n-2}
\right|^2+...  \nonumber \\
&&~~~~~~~~~~~~~~~~~~~~~~~~~~~~~~~~~~~~~~~~~~~~~~~~~~~~~~~~~~~~~~~~
...+\sum_{k_1k_2...k_n}\left|f_{0k_1}(t)f_{0k_2}(t)...f_{0k_n}(t)
\right|^2=1\;,
\label{z11a}
\end{eqnarray}
an identity of the type we are looking for, compare with Eq. (\ref{z6}). This
identity is what we
call in the present paper as {\it sum rules}. Other similar
identities can be established for other related processes. Our {\it
sum rules} are very different from the quantum mechanical sum rules
\cite{jackiw} where some identities are established by the only use
of the algebra between canonically conjugated variables. In the
present case we used the same name since our {\it sum rules} can be
established by the only use of the algebra of the matrix elements
that diagonalize the Hamiltonian.

From Eq. (\ref{z11a}) we can identify the respective probabilities
associated with all the other possible processes, thus, by comparing
it with the Eq. (\ref{z6}) we obtain
\begin{equation}
{\cal P}_{n0...0}^{(n-i)1_{k_1}1_{k_2}...1_{k_i}}(t)=
\frac{n!}{i!(n-i)!} \left|f_{0k_1}(t)f_{0k_2}(t)... f_{0k_i}(t)
[f_{00}(t)]^{n-i}\right|^2\;, ~~~i=1,2,...,n\;,  \label{z12}
\end{equation}
from which we can also obtain the corresponding probability
amplitudes which are given as
\begin{equation}
{\cal A}_{n0...0}^{(n-i)1_{k_1}1_{k_2}...1_{k_i}}(t)= \sqrt{\frac{n!}{
i!(n-i)!}}f_{0k_1}(t)f_{0k_2}(t)...f_{0k_i}(t)
[f_{00}(t)]^{n-i}\;,~~~i=1,2,...,n\;.
\label{z13}
\end{equation}
As a check, that expression above is a valid expression,
we set $i=n$ and $k_1=k_2=...=k_n=k$ in Eq. (\ref{z13}) and we
obtain Eq. (\ref{z5}) which is the probability amplitude to the emission
of $n$ field quanta of frequencies $\omega_k$ as it must be.

From Eq. (\ref{z12}) we can also obtain the probability of the dressed
oscillator to decay, at the time $t$, from the $n$-th
to the $m$-th excited level by emission of $i=(n-m)$ field quanta
of arbitrary frequencies. We denote this quantity by ${\cal P}_{n\to m}(t)$.
For this end, in Eq. (\ref{z12}), we sum up over all possible values of
$k_1$, $k_2$,...,$k_i$ and using the identity
\begin{equation}
\sum_{k}|f_{0k}(t)|^2=1-|f_{00}(t)|^2\;,
\label{z13a}
\end{equation}
that is obtained from the orthogonality property of the matrix
elements $\{t_\mu^r\}$, we get
\begin{equation}
{\cal P}_{n\to m}(t)=\frac{n!}{m!(n-m)!}\left(|f_{00}(t)|^2\right)^m
\left(1-|f_{00}(t)|^2\right)^{(n-m)}
\;,~~~m=0,1,...,n;
\label{z13b}
\end{equation}
where we can note that ${\cal P}_{n\to m}(t)$ is given by a binomial distribution
\cite{reif}.

To our knowledge no similar result to Eq. (\ref{z12}) or (\ref{z13}) has been
obtained previously. On the other hand, for weak coupling $|f_{00}(t)|^2=e^{-\Gamma t}$,
we can write  Eq. (\ref{z13b}) as
\begin{equation}
{\cal P}_{n\to m}(t)=\frac{n!}{m!(n-m)!}e^{-m\Gamma t}(1-e^{-\Gamma t})^{n-m}\;,
~~~m=0,1,...,n;
\label{beckr}
\end{equation}
a result similar to the one obtained in the early days of quantum mechanics
\cite{beck} (in the notation of this reference $i=n-m$) by a method similar to the
Einstein derivation of the black-body radiation formula.

\section{Concluding remarks}
The probability amplitudes for other processes, given by Eq. (\ref{dc1}),
can be obtained by using the crossing relation present in
that equation. We can see from Eq. (\ref{dc1}) that the probability
amplitude of an initial state $  |n_0,n_1,...,n_N\rangle_d$ to be
found at time $t$ in the state $  |m_0,m_1,...,m_N\rangle_d$ is the
same as the probability amplitude of the initial state
$|m_0,m_1,...,m_N\rangle_d$ to be found at time $t$ in the state
$|n_0,n_1,...,n_N\rangle_d$. In particular this means that the
probability amplitude of emission of a field quanta is the same as
the probability amplitude of absorption of the field quanta. We have
just computed, Eq. (\ref{z13}), the probability amplitudes related to the
emission of field quanta. Using the crossing symmetry
mentioned we can compute all the probability amplitudes related to
the absorption of field quanta. The crossing relation mentioned is
different from the field theoretical crossing relations, where the
probabilities are equal and not the probability amplitudes as it is in
the present case. The reason for the occurrence of this in our
present model is because the wave function of an harmonic oscillator
is real.

Thus, we have shown that any probability amplitude associated to the
radiation processes of an harmonic oscillator in interaction with a
massless scalar field can be given in terms only of two quantities:
the probability amplitude of the particle oscillator to remain in the first
excited level $f_{00}(t)$ and the probability amplitude of the
particle oscillator to decay spontaneously by emission of a field quantum of
frequency $\omega_k$, $f_{0k}(t)$. We have also shown that the
probability for the spontaneous decay of the oscillator from one excited level
to a lowest one, by emission of arbitrary field quanta, is given by a binomial
distribution.

Finally we have to stress that in computing the probabilities associated
to the above mentioned processes,  the calculations are greatly simplified.
This can be noted particularly in computing the integral given by Eq. (\ref{tcof}).
If no dressed coordinates were introduced, the integral that would appear instead
will not contain the exponential factor $e^{-\sum_{r=0}^N\Omega_r Q_r^2}$, but other
more complicated term that will prevent us from using directly the orthogonality
properties of the Hermite polynomials. In this way no exact calculations
will be possible. Then, we believe that the use of the dressed coordinates concept will
greatly simplify the study of early works, where extensive use has been made of the
model with Hamiltonian given by Eq. (\ref{Ham}) to model different physical situations,
such as the quantum Brownian motion, decoherence and other related problems in quantum
optics.
The study of these problems, in the framework of dressed coordinates, are under
study and will be reported elsewhere.

\vspace{0.5cm}

\begin{center}
{\large {\bf Acknowledgements}}
\end{center}

We acknowledge the anonymous referee for valuable suggestions.
GFH (grant 02/09951-3) and  RC (grant 01/12611-7) thank to FAPESP
for full support. BMP thanks CNPq and FAPESP (grant 02/00222-9) por
partial support.

\end{document}